# Providing an Approach to Predicting Customer Quality in E-Commerce Social Networks Based on Big Data and Unsupervised Learning Method

**Mohammad Arab [a,*]**

[a] Department of Entrepreneurship, University of Tehran, Tehran, Iran

* Correspondence: mohammadarab@ut.ac.ir, mohabmuse@gmail.com

**Abstract:** One of the goals of every business enterprise is to increase customer loyalty. The degree of customer loyalty is called customer quality which its forecasting will affect strategic marketing practices. The purpose of this study is to predict the quality of customers of large e-commerce social networks by big data algorithms and unsupervised learning. For this purpose, a graph-based social network analysis framework was used for community detection in the Stanford Network Analysis Platform (SNAP). Then in the found communities, the quality of customers was predicted. The results showed that various visits with an impact of 37.13% can have the greatest impact on customer quality and the order of impact of other parameters were from highest to lowest: number of frequent customer visits (28.56%), role in social networks (28.37%), Indirect transactions (26.74%), activity days (25.62%) and customer social network size (25.06%).

## 1. Introduction

Consumer behaviour has been one of the most important topics for marketing researchers in recent decades. Diversity of consumer behaviour is due to the diversity of factors affecting a person's behaviour and motivation to buy. One of the most important consumer behaviours that has positive and long-term consequences is his loyalty which the degree of loyalty is called customer quality (**Ballestar MT, 2019**). The characteristics of a customer loyalty are: Repeatedly visits to use the desired services, Uses the desired goods and services comprehensively, it tries to attract the attention of others and refer them to use goods and services and has little sensitivity in using competitors' goods and services (**KTJJoSm, 2011**). Given such characteristics, companies and organizations will make great efforts to increase their loyal customers in any way. The capabilities of today's modern customers in attracting information along with sharing ideas and experiences about different products, brands, organizations and shopping websites greatly influence their purchasing decisions and intentions. The success or failure of a product or service will not go unnoticed by consumers. (**Ho JY, 2010**). Many researchers have cited a company's ability to motivate its customers to engage in positive word-of-mouth behaviour as a strategic opportunity to achieve other loyalty outcomes. They suggest that potential customers are more likely to buy products when exposed to positive word of mouth (**Roy SK, 2014**). The digital transformation of companies has a major impact on all areas of business, especially marketing. Social media is a place with an unstable audience and loyalty is at its lowest. In social networks, it increases the loyalty of existing customers through recommendations. This strategy is one of the main areas of success of this business model, because customers who follow the next recommendations are more active and therefore more profitable and loyal to the brand. However, new users who visit these sites vary considerably in the number of transactions they make on the site. By data mining methods in big data, it is possible to predict the behaviour of such customers and cause commercial companies to change their investment method based on such predictions (**Ballestar MT, 2019**). One of the current researches in the field of e-commerce is smart e-commerce systems (SESs) (**Song Z, 2019**). Given that, the purpose of this study is to predict the quality of customers of large e-commerce social networks by big data algorithms and unsupervised learning algorithms. To achieve this goal and according to the Social Network Analysis (SNA) framework, it is necessary to first do Community Detection (**Wang T, 2011**). First, we detect social networking community by big data algorithms and unsupervised learning algorithms, and then we predict the quality of customers in the detected community.

## 2- Research background

### 2-1- Introducing common Community Detection methods in social networks

The appropriate Community Detection algorithm should be able to detect overlap, ability in directional graph, ability in weighted graph, dependence on parameter and also minimum time complexity (**Wang Y, 2018**). The social network analysis framework, especially Community Detection, can benefit from information management techniques by using an appropriate information model. There are countless information models that have varying degrees of progress, but cannot provide adequate support for Social Network Analysis (**Wang T, 2011**). The Internal Density method considers the association as a group of nodes with more adhesion between them, meaning that part of the graph has significantly more edges between its nodes than the rest of the graph. The number of edges within a community should be greater than the number of edges that connect that association to other parts of the graph. For this purpose, many algorithms have been proposed, which can be referred to as modularity algorithms (**Clauset A, 2004**), local density (**Schaeffer SE, 2005**). The modularity algorithm is based on the premise that a graph has a community structure when it differs from a random graph. Group detection based on modularity algorithm is a kind of hierarchical clustering from bottom to top approach (**Clauset A, 2004**).

Local density algorithms are of particular importance because they are local and do not require the entire graph structure. Due to the fact that in these algorithms, the addition of nodes is updated at each stage, they are divided into two static categories such as Nibble and dynamic algorithms (**YUKSEL, 2013**). The Bridge Detection method looks at a complex network as components that are connected by bridges. If we can identify these bridges, by removing them, we will reach the components that are the desired community. The most comprehensive bridge detection algorithm has been proposed by Newman and Girvan (**Girvan M, 2002**). One of the problems of this algorithm is having a time order above O(n³) which results from recalculations after detecting the first bridge. Algorithms have been proposed to improve the Newman and Girvan algorithms, the CONGA[1] algorithm operates locally and does not need to be updated throughout the graph (**Gregory S, 2008**). In the Diffusion method, community is a set of nodes that are affected within a network by the spread of a common feature or specific information. The distribution of degrees in a community is not a uniform distribution, so that the closer we get to the centre of the community boundary, the more important vertices (ie higher degrees) we encounter and the higher degree of a vertex in the graph relative to its neighbour can to some extent indicate the more important role of that vertex (**Raghavan UN, 2007**). In the Closeness method, the community is considered as a set of nodes in a group that can reach each other easily and with a small number of steps. Nodes within the community are physically close together, while nodes outside the community are further apart. The algorithms that fall into this category are generalized from random walking. The Walktrap algorithm adopts the same random walking behaviour (**Pons P, 2005**). In the Structural method, a community can be considered as a structure of immutable edges that are a combination of smaller networks with a motif structure (i.e., the existence of sub-components with a predetermined structure). Well-known algorithms in this field include K-Cliques (**Palla G, 2005**), S-Plexes Numeration (**Komusiewicz C, 2009**) and Bi-Clique (**Lehmann S, 2008**).

Link Clustering is another method of Community Detection. In this type of algorithm, instead of clustering the nodes, they cluster the edges and consider the community as a set of edges that have a common characteristic. One of the algorithms of this method is Modularity Link, which considers nodes as the same number of edges and edges as ninety common nodes, and then identifies and determines the community using a simple modularity algorithm for this graph (**Evans T, 2009**). Hierarchical Link Clustering (HLC) algorithm defines the similarity of edges based on Jaccard coefficient and accordingly uses a hierarchical clustering method on the graph (**Ahn Y-Y, 2010**). The Feature Spacing method uses hidden nodes for modelling. This algorithm brings the modelled graph closer to the main graph as much as possible and then connects each hidden node to the main node in that modelled graph. It then connects the two hidden nodes that are connected to the same main node (**Fayyad U, 1996**). Algorithms such as PMM (**Tang L, 2009**) and Infinite Relational (**Kemp C, 2006**) and AutoPart (**Chakrabarti D, 2004**) and Context-Specific Cluster Tree (**Papadimitriou S, 2008**) also fall into this category. Context-specific Cluster Tree method means finding communities at middle levels without user intervention and based on theoretical principles of information. Because communities are at the middle levels of the graph and we want to obtain communities without user intervention and based on theoretical principles of information, the Context-specific Cluster Tree algorithm is the best choice for Community Detection.

### 2-2- Unsupervised learning algorithms and K-means

Clustering is one of the unsupervised learning algorithms and a common technique for data analysis (**Ngai EW, 2009**). The main purpose of K-means is to cluster the data so that there is the least inter-cluster similarity and the most intra-cluster similarity. Intra-cluster means the distance between the two centres of the cluster and means inside the cluster is the distance between the centre of the cluster and the points assigned to that cluster. In this algorithm, the value of K is the input. A number of points are randomly selected and an initial clustering is performed and the clusters are updated based on the proximity of the distance. After the update, the centre points of the clusters are changed using averaging, and this cycle is repeated. This process continues until there is no more change in the clusters. The most popular distance calculation criteria in the clustering method are Euclidean distance and Hemming distance (**Shah N, 2014**). The basis of the Context-Specific Cluster Tree method (**Papadimitriou S, 2008**) is clustering, but this method is not designed for big data and machine learning systems. Therefore, based on the big data approach, we review the work of various researchers to select a suitable method for clustering. In (**Silva JA, 2013**) researched dimensional reduction methods, first calculating the correlation matrix from the records in the main data set and then extracting specific values according to that matrix. These eigenvalues will then be multiplied by the initial matrix, and finally the initial space property will be mapped to a new space, and these properties will be sorted, and clustering will occur based on these properties. In (**Anchalia PP, 2013**) proposed a non-automated-parallel distribution algorithm. This algorithm is complex in terms of development but has been greatly improved in terms of speed and performance. They used the DBDC[2] algorithm, which is a volume-based and distribution algorithm. The distribution algorithm is based on the density of points. This method uses compression clusters and is used when there are outliers in the data. In (**Barrachina AD, 2014**) developed techniques to improve speed and scalability. Instead of clustering the whole data, these algorithms consider samples of the data and, after clustering them, generalize the results to the whole data. Because they perform clustering algorithms on smaller pieces of data, they increase speed and are less complex. They used a macro application clustering algorithm based on random sampling called CLARANS[3]. This algorithm reduces the complexity and execution time required for the total number of data objects and has $O(n^2)$ complexity. In (**Yuwono M, 2014**) used a method to optimize the K-means algorithm to reduce dependencies and thus improve function performance. This change in performance is based on improving the speed of the algorithm. In this method, the steps after the implementation of the MapReduce function are considered in order to integrate the results optimally (**Yuwono M, 2014**). In (**Chadha A, 2014** ), introduced a method that eliminates the dependence on the value of K and does not need to provide this value as input to the algorithm. Two clusters are initially produced. Then the repetitions are done based on the classical K-means algorithm. On each of these clusters, the operation of calculating the new data centre and calculating the distances is performed, provided that there is no drop point in any of the clusters (**Li D, 2015**). In proposed a way to optimize the K-means algorithm for implementation through the MapReduce function. This method has made changes to the initial selection of cluster centres to improve the performance of the function. In this method, the distance between the sample points is calculated based on Euclidean formula and is stored in a matrix called D. Set A is initialized with the shortest distances, and set C is the centre of the clusters. Then the next steps of the algorithm are performed with matrix D. The second smallest distance in each cluster is compared to the average distance between all points. If the distance is less than the average value, it is added to set A as the smallest distance. It goes to the third point in terms of the smallest value in the D matrix, and if it is greater than the mean value, it is added to C as the centre of the cluster. In this method, there is no need to recalculate the distances between the sample points. Repeat this operation until the number of C members is equal to the number of clusters or K. (**Dierckens KE, 2017**) In presented a study called Data Science and Engineering Solutions for Fast K-means Big Data Clustering. In this study, a method called CluDataSE[4] was designed. This method was based on exploratory prototype and is a combination of clustering and temporal-spatial density methods and is faster than similar methods.

---

[1] Cluster-Overlap Newman Girvan Algorithm

**2-3- Quality of customers in e-commerce social networks**

The quality of e-services is widely defined, which includes all customer interactions in online services and sales (**Ojasalo JJIJoA, 2010**).Service quality is one of the variables of marketing science that is significantly influenced in the development of information and communication technology (**Ha HY, 2008**), the quality of electronic services of online service users from the beginning to the end of the transaction, including information search, website navigation, order processing, Interact with customer service, shipping, return policies. Efficiency in evaluating the quality of online services is of major importance (**Sahadev S, 2008**). Quality of service increases customer trust in products and companies (**MJJoR., 2016**) and trust occurs when one party involved in the exchange process believes in the other party's reliability and honesty (**Morgan RM, 1994**). Some important elements of trust are: past experiences and actions, trustworthy personality, willingness to put oneself at risk, feeling safe and confident (**SJJoSM., 2002**). By providing good services, consumers tend to feel satisfied (**Giovanis AN, 2014**). Conversely, consumers feel frustrated if a service provider provides poor service (**WJAMJ., 2013**). In fact, consumer confidence in a product creates consumer loyalty.

---

[2] Density Based Distributed Clustering
[3] Clustering Large Applications based on RANdomized Search
[4] k-means clustering data science and engineering

Consumers will talk and interact with others about the product and interact with them (**HJJoBM., 2004**). Satisfied consumers will make frequent purchases. Therefore, to build long-term customer loyalty, service providers must ensure that they not only satisfy the customer but also make them feel very satisfied. Customer loyalty is strongly influenced by customer experience and satisfaction (**Jin B, 2008**). The effect of e-trust and e-satisfaction using the sales system in online transactions shows that there is a relationship between e-service quality, e-satisfaction and e-loyalty, and the higher the quality of service, the higher the level of loyalty (**Ghane S, 2011**). Ballestar et al. Believes that the factors affecting the quality of customers in an e-commerce social network are: total direct transactions, click on direct transactions, register direct transactions, direct purchases, guidance for direct transactions, indirect transactions, activity days, role in the network social media, the size of the customer's social network, the number of frequent customer visits, various visits and conversion rates (**Ballestar MT, 2019**).

### 3- Research Methodology

In the middle region of a large graph, an arbitrary node is connected to a large number of nodes, while it is far away from a number of other nodes. If there are many paths from a to b, then from a, b is similar. On the other hand, if such a condition exists for b, then from b, a is similar. Thus, a and b are similar, if and only if there are paths between a and b as well as between b and a (**Papadimitriou S, 2008**). In fact, "two nodes in a graph are the same if and only if they are connected by several paths." The greater the number of paths between two nodes in a graph, it can be concluded that the two nodes will be more similar (**Rooshenas A, 2014**).

To achieve similarity, we act according to the following definitions and relationships:

Feature Definition: A feature denoted by the expression $G = (V, E, C)$ in which V is a set of nodes, $< a, b > \in E$ is a relation from a to b, of which a and b are members of network nodes. C is the actual class of each node. The overall size of the graph m is assumed.

Definition of similarity function: The similarity function of graph G receives two nodes a and b as input and then calculates and returns the value of similarity between two nodes a and b: $d(a, b) \to Similarity(G, a, b)$.

Graph clustering: Graph clustering means dividing a graph G into k separate parts and mathematically it can be expressed in the style of $G_i = (V_i, E_i, C)$ in which:

$$V = \bigcup_{i=1}^{k} (V_i) \ and \ V_i \cap V_j = \emptyset \ for \ any \ i \neq j \tag{1}$$

An appropriate clustering algorithm should work to increase the value of Eq. (2) in the case of x = y and decrease the value of the relation in the case of x ≠ y.

$$Min \sum_{i=1}^{k} \sum_{x_j \in C_i} \|x_j - c_i\|^2 \tag{2}$$

Which is a set of data from n data points $(x_1, \dots x_n)$ and the number of k clusters $(C_1, \dots, C_k), k \leq n$. Which according to Equation (1) will have:

$$\sum_{i \in G_x} \sum_{j \in G_y} d(v_i, v_j) \tag{3}$$

Definition of Access Value: Suppose P is the transfer probability cube, N * N * p for the graph G. The access value from a to b is calculated as follows:

$$H(a, b) = w_1 P_{a,b}^1 + \cdots + w_p P_{a,b}^p + \cdots + w_{n-2} P_{a,b}^{n-2} \tag{4}$$

And for every a and b we will have:

$$H_k(a, b) = w_1 P_{a,b}^1 + \cdots + w_p P_{a,b}^k \ 1 \leq k \leq n - 2 \tag{5}$$

Where $w_i$ is the weight of all paths of length i and $P_{a,b}^p$ is the probability of going from a to b to length p and is equal to the number of paths of length p from a to b $(K_p(a, b))$, divide by the total number of paths of length p starting from a, $(K_p(a, x))$.

$$P_{a,b}^p = \frac{K_p(a, b)}{\sum_{\forall x \in G-(a)} K_p(a, x)} \tag{6}$$

Therefore, in order to obtain meaningful results, it is necessary that the weight of short paths is more than the weight of longer paths. The best way to get the similarity of the two nodes a and b in the given graph is to normalize $H_k(a,b)$ with respect to $H_{Max}$ and $H_{Min}$. Therefore, the Feature Spacing will be calculated with the following expression:

$$Feature\ Spacing_k = \frac{H_k(a,b) - H_{Min}}{H_{Max} - H_{Min}} \tag{8}$$

Based on Equations 1 to 7, the Feature Spacing can be obtained by extracting different paths between the two nodes in the given graph. The algorithm in Figure 1 shows the recursive procedure for calculating the Feature Spacing in the given graph G.

| | |
|---|---|
| 1 | $PathQueue \leftarrow New\ Queue$ |
| 2 | $AllValidPath \leftarrow New\ String[m]$ |
| 3 | $for\ all\ (a,b) \in G\ do$ |
| 4 | $\quad if\ (a,b) \in E\ then$ |
| 5 | $\quad\quad AllValidPath[a] \leftarrow AllValidPath[a] + (a,b)$ |
| 6 | $\quad\quad PathQueue.insert((a,b))$ |
| 7 | $\quad end\ if$ |
| 8 | $end\ for$ |
| 9 | $repeat$ |
| 10 | $\quad tmpPath \leftarrow PathQueue.Remove()$ |
| 11 | $\quad firstNode \leftarrow tmpPath.getFirstNode\ ()$ |
| 12 | $\quad lastNode \leftarrow tmpPath.getLastNode()$ |
| 13 | $\quad for\ all\ (lastNode,b) \in G\ do$ |
| 14 | $\quad\quad AllValidPath\ [firstNode] \leftarrow AllValidPath\ [firstNode] + (tmpPath,b)$ |
| 15 | $\quad\quad PathQueue \leftarrow PathQueue + (tmpPath,b)$ |
| 16 | $\quad end\ for$ |
| 17 | $until\ (tmpPath.lenght() \leq p)$ |
| 18 | $for\ all\ (a,b) \in G\ do$ |
| 19 | $\quad H(a,b) \leftarrow 0$ |
| 20 | $\quad Feature\_Spacing(a,b) \leftarrow 0$ |
| 21 | $end\ for$ |
| 22 | $for\ all\ a \in G\ do$ |
| 23 | $\quad for\ (l=1;l \leq p;l++)\ do$ |
| 24 | $\quad\quad for\ all\ b \in G - (a)\ do$ |
| 25 | $\quad\quad\quad Path \leftarrow \frac{|AllValidPath\ [a].getPath(b,l)|}{|llValidPath\ [a].getPath(null,l)|}$ |
| 26 | $\quad\quad\quad H(a,b) \leftarrow H(a,b) + Path$ |
| 27 | $\quad\quad end\ for$ |
| 28 | $\quad end\ for$ |
| 29 | $end\ for$ |
| 30 | $for\ all\ (a,b) \in G\ do$ |
| 31 | $\quad Feature\_Spacing(a,b) \leftarrow \frac{H(a,b)-Min(H)}{Max(H)-Min(H)}$ |
| 32 | $end\ for$ |

**Figure.1** The proposed recursive algorithm for calculating the Feature Spacing

The algorithm has four basic steps, initialization, route extraction, access value calculation and Feature Spacing calculation. In the first step, an empty queue is created to hold the processed paths as well as an array of observed paths for each node (line 2). *PathQueue* and *AllValidPath* are then quantified by adding all the edges in the G input graph. In the second step, all paths of less than or equal to p in length are extracted in an efficient method. Lines 17 and later show the return method for extracting paths, in which *tmpPath* contains the first path in *PathQueue*. Output nodes, the last node in *tmpPath* is added to *tmpPath* itself until new paths are created. This adds new paths to *AllValidPath* and *tmpPath*. The third step calculates the access value for each pair of nodes according to Equations 5 to 6. Finally, Feature Spacing will be calculated in the last step based on Formula 7. The access value calculation step is the heaviest operation in the algorithm of Figure 1, and the H value is calculated for different paths between each node. The for loop in line 22 is loaded a maximum of m times, the for loop in line 23 is run a maximum of p times, which is the length of the paths selected by the user. Also, the for loop is executed on line 24 up to m times. Therefore, the temporal complexity of the characteristic distance algorithm is equal to $O(m^2 s)$, which in the worst case would be equal to $O(m^4)$. The spatial complexity of *AllValidPath* as well as H is equal to $O(m)$ and $O(m^2)$, respectively (**YUKSEL, 2013**).

In the larger data set, the more time frame it takes to run step 2 because it visits each data item and performs calculations on it. The question here is: do we need to visit the entire data space? In fact, most data objects belonging to clusters or clusters whose centres move slowly should not be affected by the movement. They will remain part of the same cluster during the next iteration, and fewer points will be affected by movement.

The ability to identify which data subjects are affected by motion allows us to perform an important function in the implementation of step number two. To prevent such calculations, we can plot all the elements of our data at wider distances as shown in the algorithm in Figure 2. This algorithm is based on the research of Dirkens et al. (**Dierckens KE, 2017**). The group points of the algorithm in Figure 2 are such that instead of visiting each data element and controlling it against a distance close to its edge and boundary, it executes it for the whole group. The key to all optimization is performance compatibility. Wide compatibility has a significant effect on optimization. If the value of the width value is low, then the number of intervals increases and the workload for checking and controlling and updating the intervals in each iteration can be significantly increased. In each iteration, the composition and arrangement of the clusters is exactly the same as the standard k-means used (**Farivar R, 2008**).

---

*1. Define constant WIDTH*
*2. Define intervals li = $l_i$ \* WIDTH,(i+1) \* WIDTH) and tag them with value i \* WIDTH*
*3. Mark the entire data set to be visited*
*4. For each point to be visited*
*5. Compute e = min($d_{pci}$ - $d_{pcw}$) where $C_w$ is the centre of the winner (closest) cluster and $C_l$, l=1..k,l≠w stands for all other centroids*
*6. Map all points with i \* WIDTH < e < (i+1) \* WIDTH to interval i \* WIDTH where i is a positive integer*
*7. Compute new centroids Cj,where j=1..k and their maximum deviation D = max(|CjCj'|)*
*8. Update Ij's tag by subtracting 2 \* D (points owned by this interval got closer to the edge by 2 \* D)*
*9. Pick up all points inside intervals whose tag is less or equal to 0, and go to 4 to revisit them*

**Figure.2** Proposed clustering algorithm based on research (**Dierckens KE, 2017**)

---

According to the procedure, the Community Detection method should be performed by the proposed Feature Spacing algorithm. This algorithm uses big data processing for clustering. Finally, customer quality prediction in found communities is calculated by an unsupervised learning algorithm. The implementation method is shown in Figure 3.

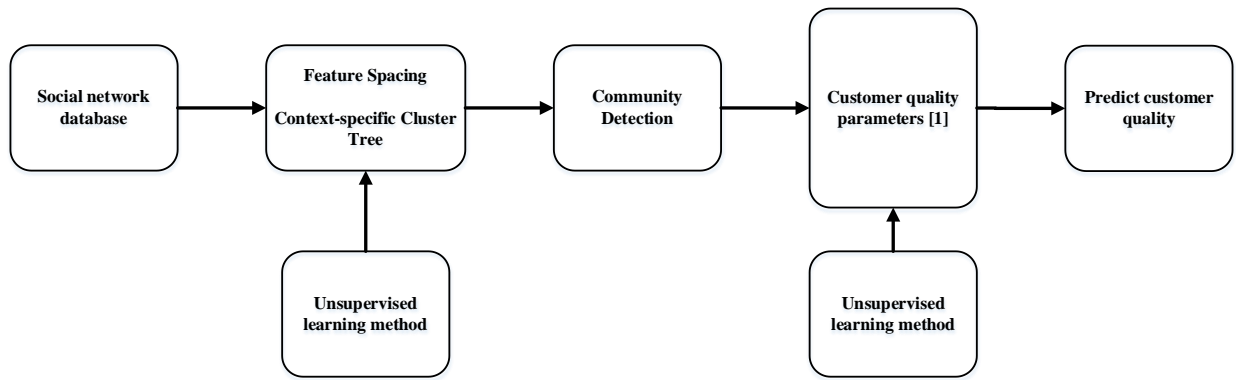

**Figure.3** Research implementation process

## 4- Results

This study used data from the Stanford Network Analysis Platform (SNAP). SNAP has a very large library and database, including millions of nodes and edges, to extract graphs from popular social networks such as Facebook and Twitter. The SNAP library is based on C ++. The software and database can be downloaded from the SNAP website[5].

---

[5] https://snap.stanford.edu/data/

We cluster the social network graph using the K-means method. We then identify communities by Feature Spacing, focusing on the middle layers. Finally, by clustering method, we predict the quality of customers based on the parameters (**Ballestar MT, 2019**). These parameters were: total direct transactions, click on direct transactions, register direct transactions, direct purchases, guidance for direct transactions, indirect transactions, business days, role in the network social media, the size of the customer's social network, the number of frequent customer visits, various visits and conversion rates (**Ballestar MT, 2019**).

The installation and execution environment of the algorithms was Ubuntu 18.04 and Intel (R) Core (TM) i7-75000 CPU @ 2.70 GHz and 16 Gb DDR3. The programming language of this research is Python. After downloading the latest version of SNAP, we will install it. Then, by cncom.py, we will connect to the database and download the necessary information shown in Table 1.

**Table.1.** Information of selected communities from SNAP (**Yang J, 2015**)

| Social network name | Number of nodes | Number of edges | Number of communities |
|---|---|---|---|
| com-LiveJournal | 3,997,962 | 34,681,189 | 287,512 |
| com-Friendster | 65,608,366 | 1,806,067,135 | 957,154 |
| com-Orkut | 3,072,441 | 117,185,083 | 6,288,363 |
| com-YouTube | 1,134,890 | 2,987,624 | 8,385 |
| com-DBLP | 317,080 | 1,049,866 | 13,477 |
| com-Amazon | 334,863 | 925,872 | 75,149 |
| email-Eu-core | 1,005 | 25,571 | 42 |
| wiki-topcats | 1,791,489 | 28,511,807 | 17,364 |

For Community Detection, the parameters: Friends Network, Like Network, Like Comment Network, Activity Recent History, Activity types, Activity Weekly Distribution, have been used (**Chang VJTF, 2018**).

Table 1 social networks do not have to include all six parameters (**Chang VJTF, 2018**). Therefore, the number of communities identified may not match the actual number of communities in the SNAP database. Table 2 shows the implementation of the algorithm for the eight social networks in the SNAP database. This table shows the number of communities found with the proposed algorithm and the amount of error. The average error of the number of communities found for eight social networks is 9.84%. The main cause of discrepancy can be related to the information available in the networks. For example, in com-LiveJournal, which has 287512 communities, there is a 4.19% error, but in com-YouTube, which has only 8385 forums, there is a 17.35% error. Which indicates that the communication of the six factors was less in the com-YouTube network. Figure 4 is a diagram drawn based on Table 2.

**Table.2.** Number of communities found with clustering and Feature Spacing

| Social network name | Number of communities in SNAP | Number of communities found | Error (percentage) |
|---|---|---|---|
| com-LiveJournal | 287,512 | 275,451 | 4.19 |
| com-Friendster | 957,154 | 843,692 | 11.85 |
| com-Orkut | 6,288,363 | 5,459,713 | 13.18 |
| com-YouTube | 8,385 | 6,930 | 17.35 |
| com-DBLP | 13,477 | 12,572 | 6.72 |
| com-Amazon | 75,149 | 70,349 | 6.39 |
| email-Eu-core | 42 | 38 | 9.52 |
| wiki-topcats | 17,364 | 15,707 | 9.54 |

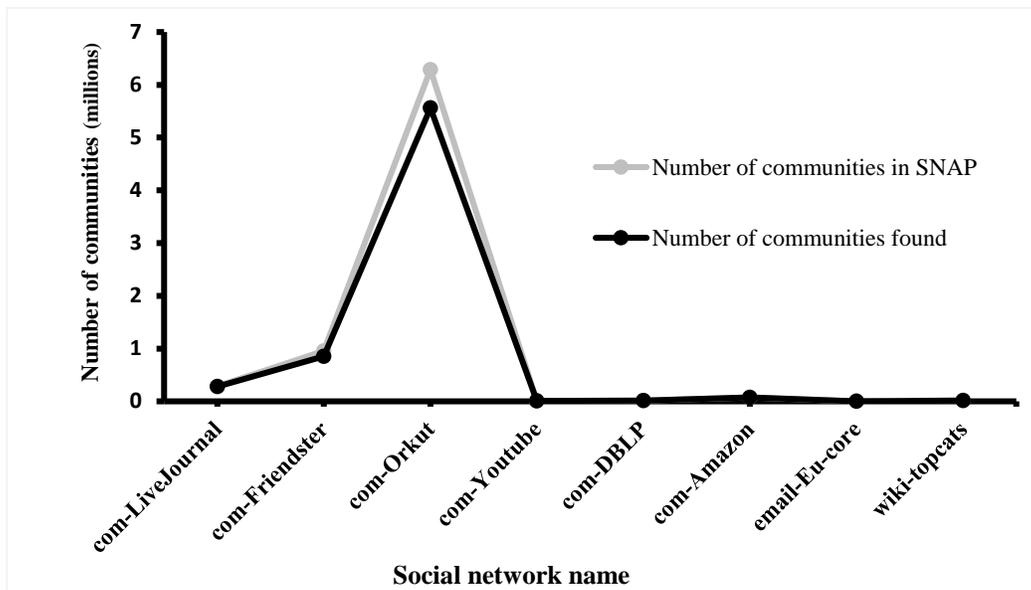

**Figure.4** Comparison of the number of communities found and real for eight social networks

After finding communities, the com-Orkut social network was selected as a candidate to predict the quality of customers. This social network included for factors affecting customer quality. These parameters include: indirect transactions, activity days, role in social networks, customer social network size, number of frequent customer visits, various visits. After calculating the prediction, the percentage of each of these parameters was obtained, which is shown in Figure 5.

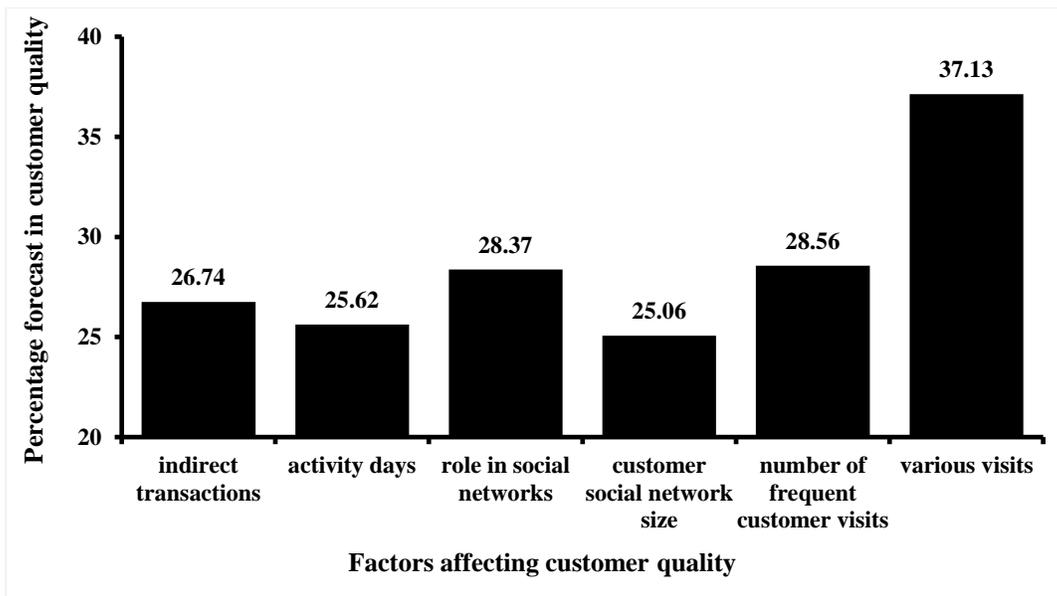

**Figure.5** Predicting the effect of factors on customer quality

## 5- Conclusion

In this research, based on the graph method, we have developed the concepts of management and analysis of large social networks. To find communities, unsupervised learning algorithms and k-means clustering, were combined with the Context-Specific Cluster Tree (**Papadimitriou S, 2008**) method. The results show that this approach has an average error of 9.82% for finding communities.

Also, the quality of customers was predicted using big-data clustering algorithm. The results showed that various visits with an impact of 37.13% can have the greatest impact on customer quality. The order of impact on customer quality from highest to lowest are: (1) various visits, (2) number of frequent customer visits, (3) role in social networks, (4) indirect transactions, (5) activity days, (6) Customer social network size.